\documentclass{article}

\usepackage[pdf]{graphviz}
\usepackage{algorithm}
\usepackage{algpseudocode}
\usepackage{hyperref}

\title{Revisiting MAB based approaches to recursive delegation}

\date{2 November 2023}
\author{Nir Oren}

\begin{document}

\maketitle

\begin{abstract}
In this paper we examine the effectiveness of several multi-arm bandit algorithms when used as a trust system to select agents to delegate tasks to. In contrast to existing work, we allow for recursive delegation to occur. That is, a task delegated to one agent can be delegated onwards by that agent, with further delegation possible until some agent finally executes the task. We show that modifications to the standard multi-arm bandit algorithms can provide improvements in performance in such recursive delegation settings.
\end{abstract}

\section{Introduction}
Open multi-agent systems (MAS) are composed of agents under different organisational control, and whose internal goals and mental states cannot be observed. In such systems, agents often have differing capabilities, and must rely on each other when pursuing their goals, making task delegation commonplace. This delegation occurs when one agent (the \emph{delegator}) requests that another (the \emph{delegatee}), execute a task. A fundamental problem faced by the delegator involves selecting the most appropriate delegatee to whom the task should be delegated, and a significant body of work centred around trust and reputation systems has examined how such a delegation decision should take place \cite{granatyr2015trust,pinyol2013computational,sievers2022modeling}.

At their heart, trust and reputation systems associate a rating with each potential delegatee, and select who to delegate a task to based on this rating. Following task execution, the rating is updated based on how well the task was completed. Different systems compute the ratings differently, for example incorporating indirect information from other agents in the system \cite{10.1145/775152.775242,teacy2012efficient}, or  utilising social and cognitive concepts as part of the computation process \cite{falcone2001social}. Trust and reputation systems can also differ in the way they select a delegatee, for example by using the rating to weigh the likelihood of selection. While trust and reputation systems seek to satisfy many properties including resistance to different types of attacks by malicious agents \cite{josang2009challenges}, at their heart, they balance the exploration of delegatee behaviour with the exploitation of high quality delagatees. Balancing exploration versus exploitation also lies at the heart of multi-arm bandit (MAB) problems \cite{kuleshov2014algorithms}, and several approaches to trust and reputation have been proposed which build on MAB algorithms \cite{vallee2014multi}.

With few exceptions \cite{afanador2019algorithms,burnett2012sub}, trust and reputation systems assume that delegation is non-recursive. That is, once a delegator passes a task onto a delegatee, it is executed by the latter. However, in many settings, a delegatee could  delegate the task on to another agent, who in turn could delegate it further. Such \emph{recursive delegation} naturally arises in the real world. For example, outsourcing is commonplace in the hydrocarbon exploration domain, where a company wishing to search for oil may task another company with finding a ship, and this company outsources crew selection to another company, etc. Failure in one part of this \emph{delegation chain} requires the first company to decide how to ascribe blame down the chain, affecting its future decisions about who to interact with.

While delegated tasks may be decomposed in different ways with various complex interrelationships (c.f., HTN planning \cite{georgievski2015htn}), in this paper we consider the recursive delegation of an atomic task. Our main contributions are as follows.

\begin{itemize}
  \item We demonstrate the importance of considering delegation in the context of the exploration versus exploitation problem.
  \item We propose and evaluate several algorithms for recursive delegation, demonstrating differences in performance between these and non-delegation aware algorithms.
\end{itemize}

The next section of the paper describes the recursive delegation problem and introduces the algorithms which we evaluate in Section \ref{sec:eval}. We discuss related and future work in Section \ref{sec:discussion} before concluding.

\section{Recursive Delegation in a MAB setting}

\begin{figure}
\begin{center}
\includegraphics[width=0.49\textwidth]{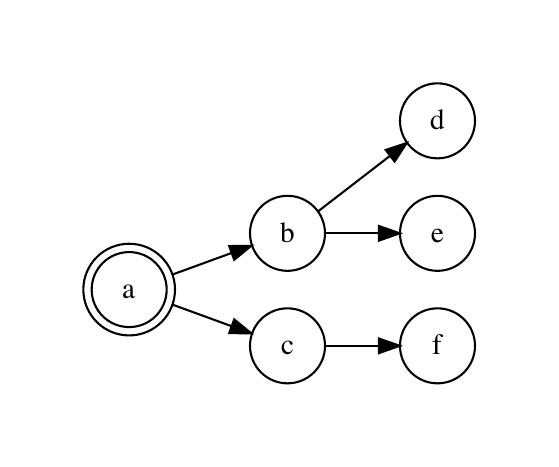}
\includegraphics[width=0.49\textwidth]{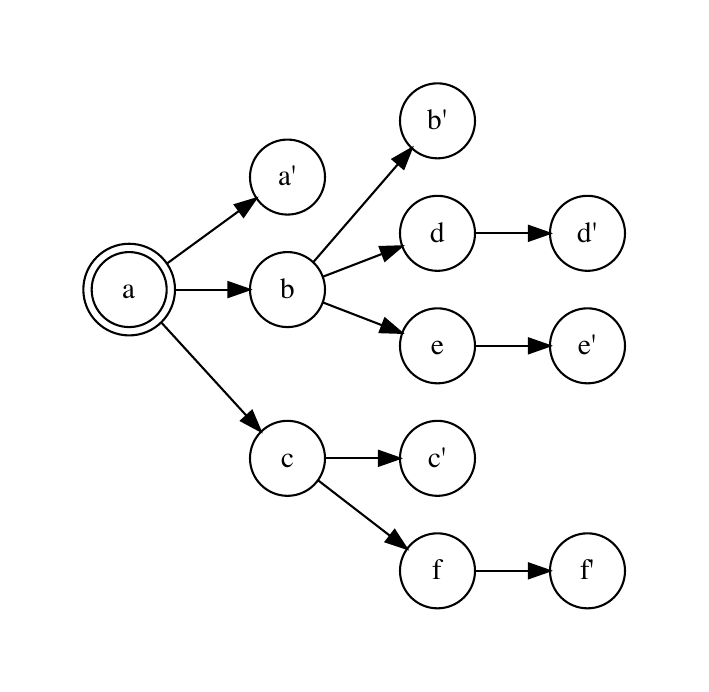}
\end{center}
\caption{\label{fig:example1}Left: a simple delegation scenario. Right: the same scenario with the addition of virtual agents.}  
\end{figure}

Consider the system of agents shown in Figure \ref{fig:example1}. Here, we assume that agent $a$ will delegate a task to either agent $b$ or agent $c$. If $b$ is delegated to, they can delegate the task to $d$ or $e$, while if $c$ is delegated to, they can delegate the task onwards to $f$. Once the task reaches $d,e$ or $f$, it will be executed.

Now assume that $a$ is unaware of the possibility of recursive delegation, and that they delegate a task to $b$, who delegates it onto $d$, with $d$ unsuccessfully completing the task. $a$ then delegates another task to $c$ (and thus onto $f$), with the task now successfully completed. In such an instance, $a$ will trust $c$ more than $b$, unaware that $b$ may next delegate onto $e$. If $e$ is more likely to successfully complete a task than $f$ (who in turn is more likely to succeed in the task than $d$), then $a$'s delegation strategy  is sub-optimal. Instead, $a$ should recognise that agent $e$ has not yet been delegated to, and its success rate is unknown, and therefore again delegate a task to $b$. It is this intuition that we attempt to capture in our MAB-based delegation algorithms. While the standard MAB based algorithms we build on consider delegation to a single node, the algorithms we propose must consider the various ways delegation can occur passing through a given agent.

Note that in this paper, we assume that only agents at the leaves of our delegation graph can execute tasks. This is not a significant restriction as any agent which can delegate or execute a task can be modified through the introduction of a virtual agent to which it can delegate as a leaf (c.f., the right side of Figure \ref{fig:example1}, where agents labelled with a prime represent such virtual agents); this virtual agent represents the agent taking an execution action.

In the rest of this section we consider several widely used MAB algorithms, namely $\epsilon$-greedy; UCB1; Bayesian (Beta) UCB; and Thompson sampling. The standard (i.e., non-delegation aware) variant of these algorithms treats each delegation action as a single instance of a MAB. Returning to Figure 2, $a$ would select between $a'$,$b$ and $c$ by treating it as a single MAB problem based on the historic success of $a'$, $b$ and $c$ and any other parameters relevant to the algorithm. If (for example) $b$ is then selected, that agent would choose to delegate to $b'$, $d$ or $e$ again based on the historic success of these agents, ignoring future delegation possibilities.

Our delegation-aware variant of the algorithms instead considers possible delegation paths. In general, loops are possible in arbitrary delegation structures, and we assume that an agent will not delegate to an agent already appearing in the delegation chain. Our primary modification of the standard MAB algorithms therefore involves changing how the expected utility of potential delegatees is calculated, which we describe next.

It is worth noting that we make one assumption across all our delegation algorithms, namely that the result of a delegation is binary; either the task succeeds, or fails. Integrating different possible levels of task success is one important avenue of future work.

\subsection{$\epsilon$-greedy}

The standard $\epsilon$-greedy algorithm selects a random agent to delegate to with likelihood $\epsilon$, and selects the agent with the highest historic success rate with likelihood $1-\epsilon$. After every interaction, the success rate of the executing agent is updated. In a recursive delegation setting, success rates are updated for all delegating agents too. 

Our adaptation of this basic algorithm involves the delegating agent considering how onwards delegation may occur, based on the expected utility of a delegation. The pseudo-code for calculating this utility is shown in Algorithm \ref{alg:epsilon-greedy-modified}. 

\begin{algorithm}
\begin{algorithmic}[1]
\Require A value $\epsilon$
\Function{Calculate-utility}{$a,chain$}
  \State $poss\_del \gets$ neighbors($a$) $\backslash chain$
  \If{$poss\_del = \emptyset$}
    \State \Return succ($a$)/(succ($a$)+fail($a$))
  \EndIf
  \State $util \gets 0$
  \State $best \gets 0$
  \ForAll{$c \in poss\_del$}
    \State $uc \gets$ \Call{Calculate-utility}{$c,chain+[a]$}
    \If{$uc>best$}
      \State $best \gets uc$
    \EndIf
    \State $util \gets \epsilon \frac{uc}{|poss\_del|}$
  \EndFor
  \State \Return $util+(1-\epsilon)best$
\EndFunction
\end{algorithmic}
\caption{\label{alg:epsilon-greedy-modified}}
\end{algorithm}

This algorithm takes an agent and the chain of delegations until that agent as input, and identifies the set of possible delegations that that agent can delegate to (assuming that an agent cannot be delegated the same task twice).  If no possible delegators exist, then the agent must execute the task, and we return the expected likelihood of successfully doing so (succ($a$) and fail($a$)  denote the number of times the agent has successfully, or unsuccessfully, executed a task respectively).

Otherwise, the utility of every possible delegation is (recursively) calculated, recognising that each such delegation is equally likely to be selected based on $\epsilon$. The utility of the optimal delegation is also calculated, allowing us to determine the net expected utility of delegating a task to agent $a$.

Once the expected utility for a delegation has been computed, we can apply the standard $\epsilon$-greedy algorithm to undertake the delegation, selecting the agent with highest (expected) utility with likelihood $1-\epsilon$, and a random agent with likelihood $\epsilon$.

\subsection{UCB}

In the delegation context, the UCB algorithm selects a delegatee based on the \emph{upper confidence bound} associated with the delegatee. With no information about the probability distribution underlying the delegatee's behaviour, this bound is computed as the mean reward received so far plus an additional term $\sqrt{\frac{2\log n}{n_0}}$ where $n$ is the total number of delegations performed so far across the system, and $n_0$ is the total number of times a task has been delegated to the delegatee under consideration.

Our adaptation of the basic UCB algorithm involves us considering the UCB of the node at which execution takes place. The UCB value of a parent node is then the maximal UCB value of all its child nodes, with a delegator delegating to the delegatee with the maximum UCB value. This calculation of the UCB value is formalised in Algorithm \ref{alg:UCB} where $C$ (in line \ref{alg:line:ucb}) is a constant which helps control the level of exploration of the algorithm.

\begin{algorithm}
\begin{algorithmic}[1]
\Function{Calculate-UCB-utility}{$a,chain$}
  \State $poss\_del \gets$ neighbors($a$) $\backslash chain$
  \If{$poss\_del = \emptyset$}
    \State \Return $\frac{succ(a)}{succ(a)+fail(a)} + C\sqrt{\frac{2\log{n}}{succ(a)+fail(a)}}$ \label{alg:line:ucb}
  \EndIf
  \State \Return $\max_{c \in poss\_del}$ \Call{Calculate-UCB-utility}{$c,chain+[a]$}
\EndFunction
\end{algorithmic}
\caption{\label{alg:UCB}}
\end{algorithm}

The difference between the mean and upper confidence bound in our system is a measure of the uncertainty in the estimate of the performance of a delegatee. By modeling this uncertainty using a Beta distribution, we can change the UCB term in line \ref{alg:line:ucb} to the standard deviation of the distribution. We refer to this variant of the UCB algorithm as the Beta-UCB algorithm. As will be shown in the next section, doing so significantly improves the performance of the algorithm.

\subsection{Thompson Sampling}

Another popular, and very effective, algorithm used in the multi-arm bandit literature is Thompson sampling. Here, the value associated with a delegatee is computed by randomly sampling from a probability distribution representing the delegatee (which in our case is the Beta distribution.

In the recursive delegation situation, we are now faced with a choice. When two possible paths reach the same task executor, we can either sample this executor twice, propagating the value upwards, or sample once, and propagate the same value up. Sampling only once provides significant performance improvements (as it facilitates caching), and was thus the approach we used. Algorithm \ref{alg:TS} summarises the utility computation under this approach.

\begin{algorithm}
\begin{algorithmic}[1]
\Function{Calculate-TS-utility}{$a,chain$}
  \State $poss\_del \gets$ neighbors($a$) $\backslash chain$
  \If{$poss\_del = \emptyset$}
    \State \Return sample from Beta($succ(a),fail(a)$)
  \EndIf
  \State \Return $\max_{c \in poss\_del}$ \Call{Calculate-TS-utility}{$c,chain+[a]$}
\EndFunction
\end{algorithmic}
\caption{\label{alg:TS}}
\end{algorithm}

\section{Evaluation} \label{sec:eval}

We evaluated the four algorithms described above against their ``standard'' counterparts in a variety of scenarios\footnote{Source code is available at \url{https://github.com/jhudsy/RecursiveDelegationMAB}} and examined the average resultant cumulative regret over multiple runs.

\begin{figure}
\includegraphics[width=\textwidth]{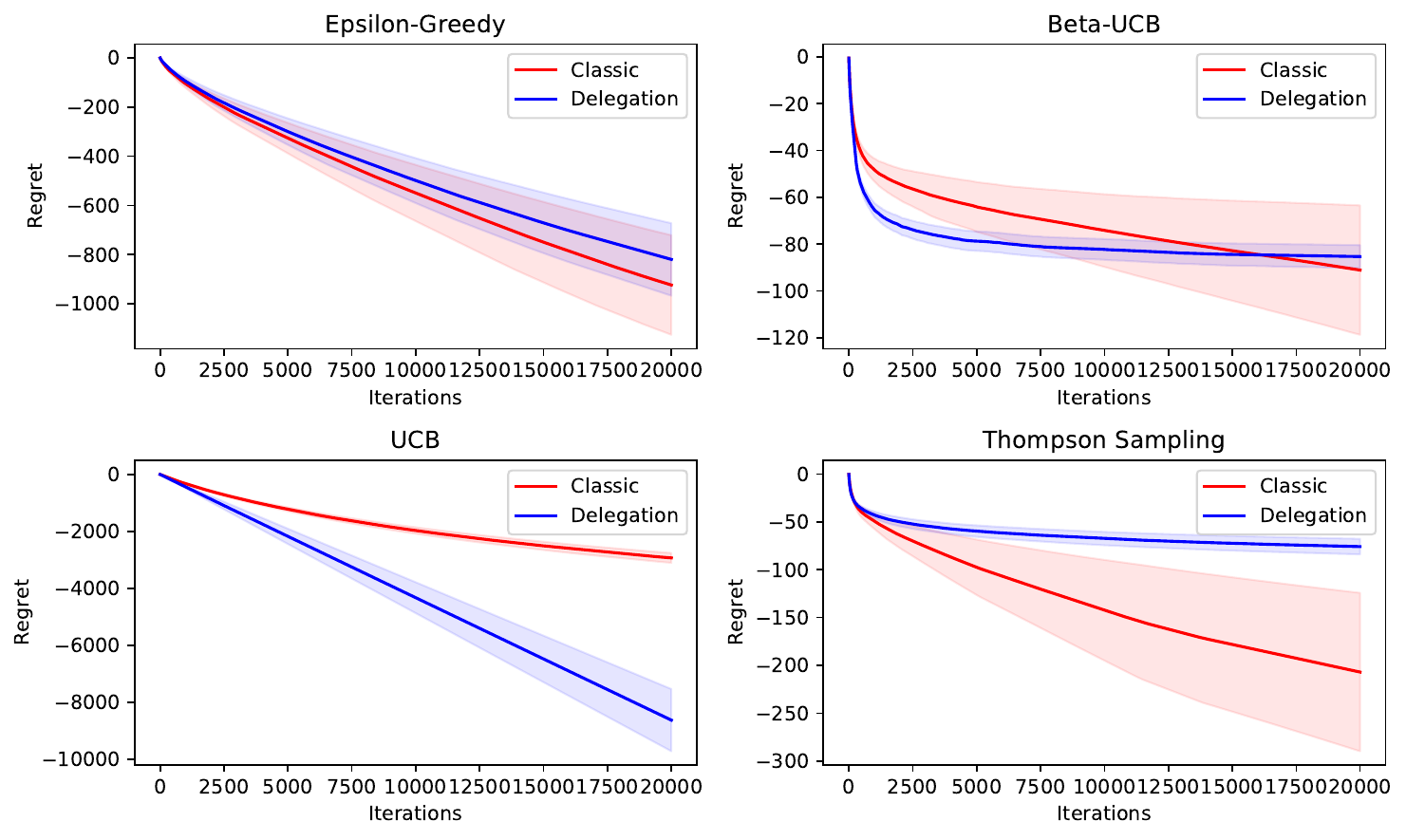}
\caption{\label{fig:eval1} Evaluation of the four algorithms with 20 agents using a binomial graph topology. Exploration parameters for both UCB algorithms were set to 3.}
\end{figure}

Figure \ref{fig:eval1} illustrates the performance of the four algorithms averaged over 100 different binomial graphs (with the probability of edge creation set to 0.3). In this figure, the UCB constant for both standard and Beta-UCB approaches was set to 3. Outwith the standard UCB algorithm, the modified, delegation-aware variants of the algorithms outperform their standard counterparts. However, for Epsilon-greedy, the performance of the two algorithms is very close, while for the Beta-UCB variant, the delegation-aware variant initially does not perform as well as it undertakes more exploration around possible delegations than the standard version of the algorithm. We believe that for the UCB case, the poor performance of the delegation-aware variant occurs as exploration dominates due to the way in which UCB is calculated.

\begin{figure}
\includegraphics[width=\textwidth]{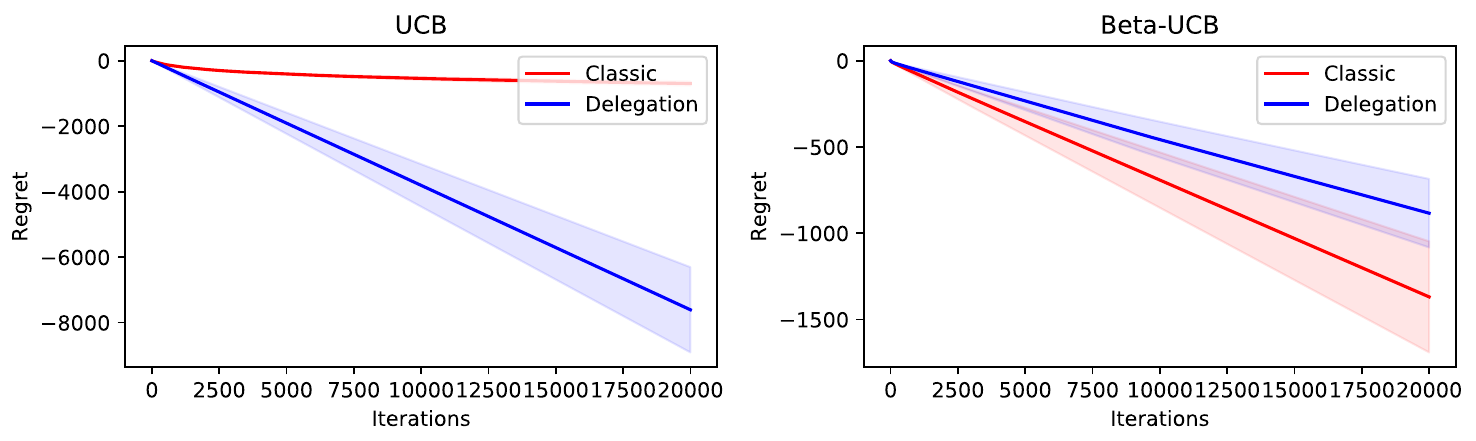}
\caption{\label{fig:eval2} Evaluation of the UCB algorithms with exploration parameters set to 1.}
\end{figure}

Figure \ref{fig:eval2} shows how the UCB variants performed when their associated constant was reduced from 3 to 1. As can be seen, the reduction in exploration allowed both standard UCB variants to reduce their cumulative regret, though the classic version of the algorithm still outperformed the delegation aware variant. Turning to Beta-UCB, we can see that while the delegation-aware variant still performed better, its regret was much worse than the case with a higher exploration constant. These results indicate that the fine-tuning of this constant is therefore of critical importance to the performance of these algorithms.

\begin{figure}
\includegraphics[width=\textwidth]{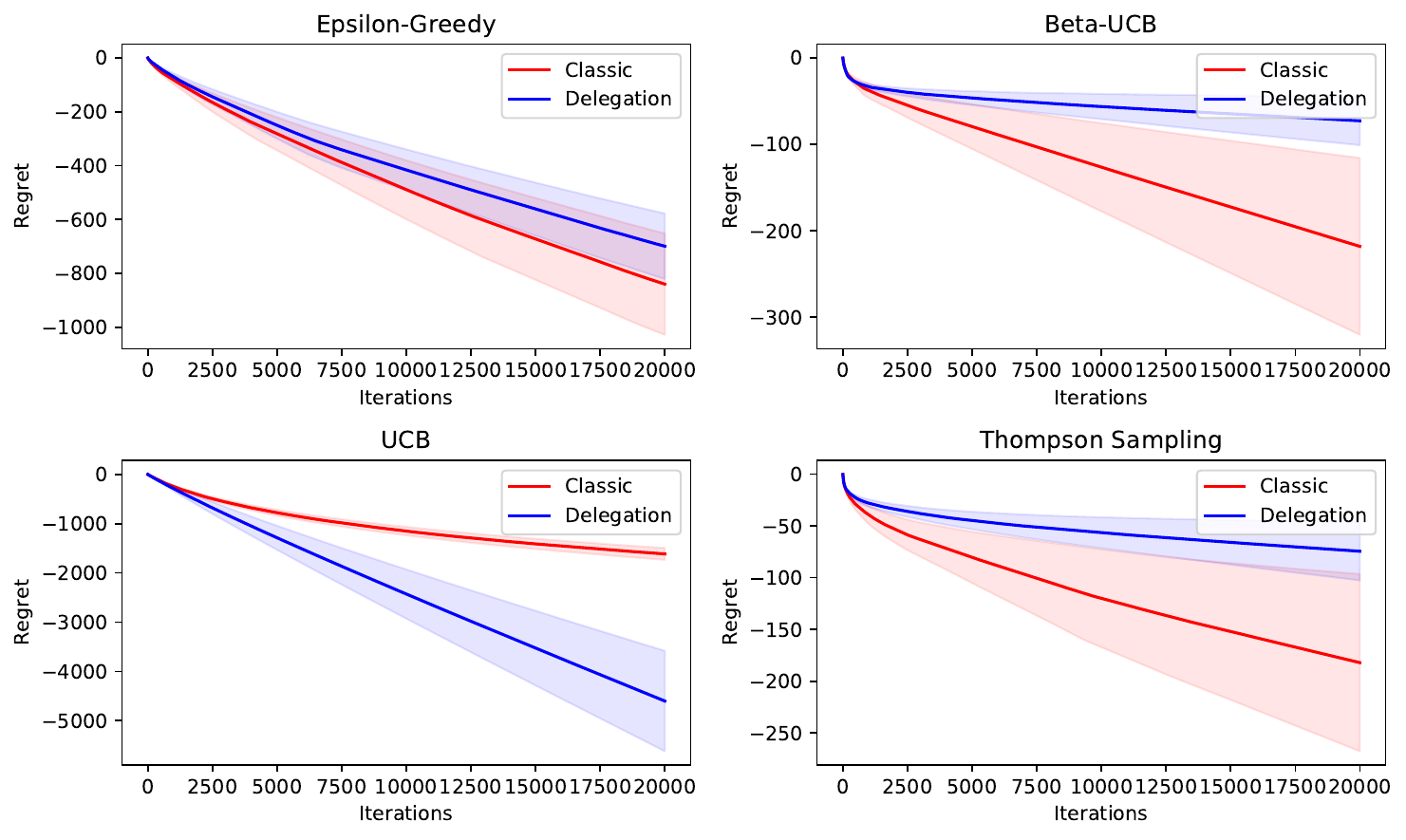}
\caption{\label{fig:eval3} Evaluation of the four algorithms with 10 agents using a binomial graph topology.}
\end{figure}

\begin{figure}
\includegraphics[width=\textwidth]{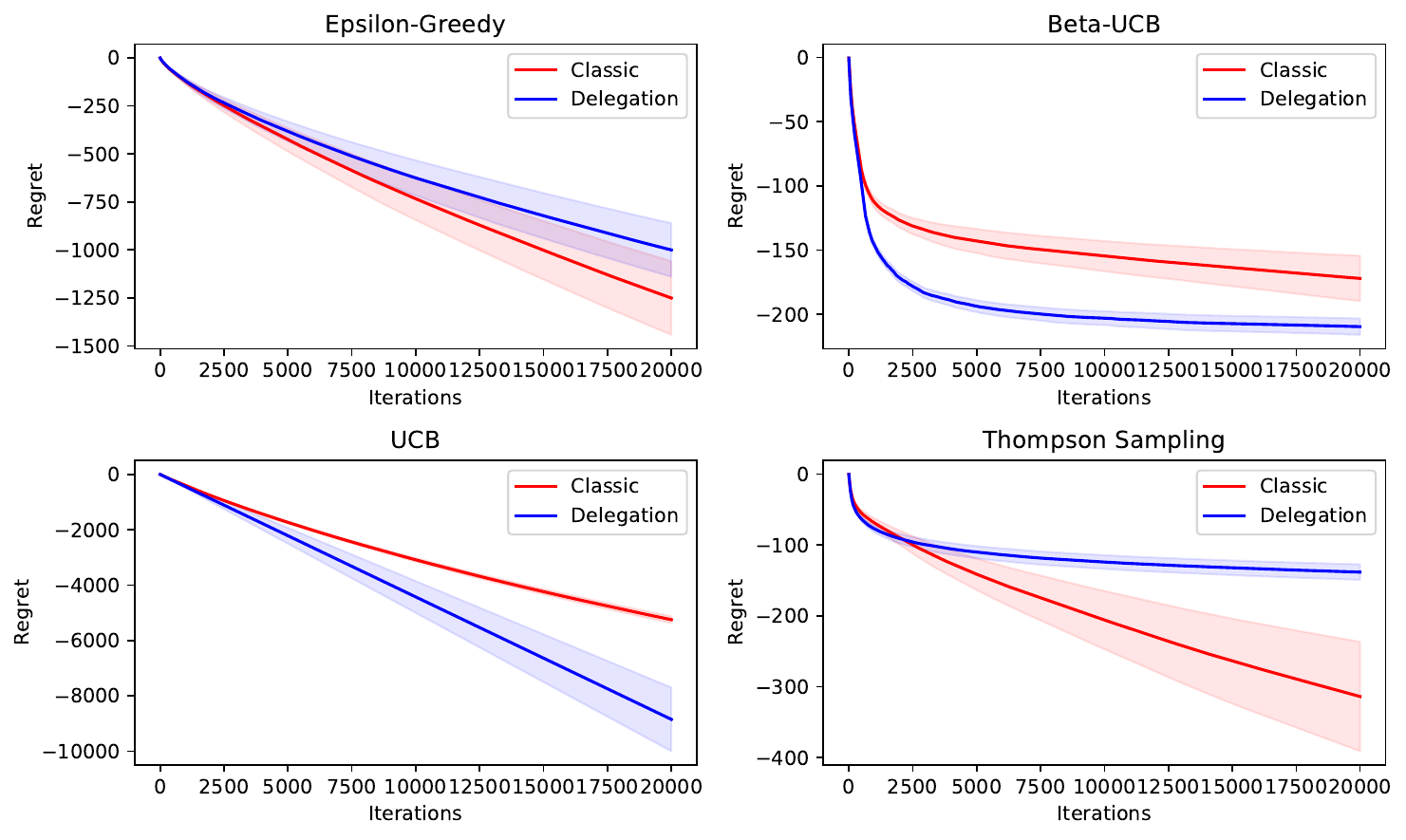}
\caption{\label{fig:eval4} Evaluation of the four algorithms with 50 agents using a binomial graph topology.}
\end{figure}

We also considered the impact of changing the number of agents in the system. An increased number of agents would potentially allow more possible delegations to occur, increasing the need for exploration. The results of this experiment are shown in Figures \ref{fig:eval3} (for 10 agents) and \ref{fig:eval4} (for 50 agents). As can be seen, with fewer agents, the performance of Beta-UCB improves. For 50 agents, after 20000 iterations, the standard version of Beta-UCB still outperforms the delegation-aware variant (though the gap is closing). For Thompson sampling, the delegation-aware version of the algorithm consistently outperforms the standard version across different numbers of agents. 

\begin{figure}
\includegraphics[width=\textwidth]{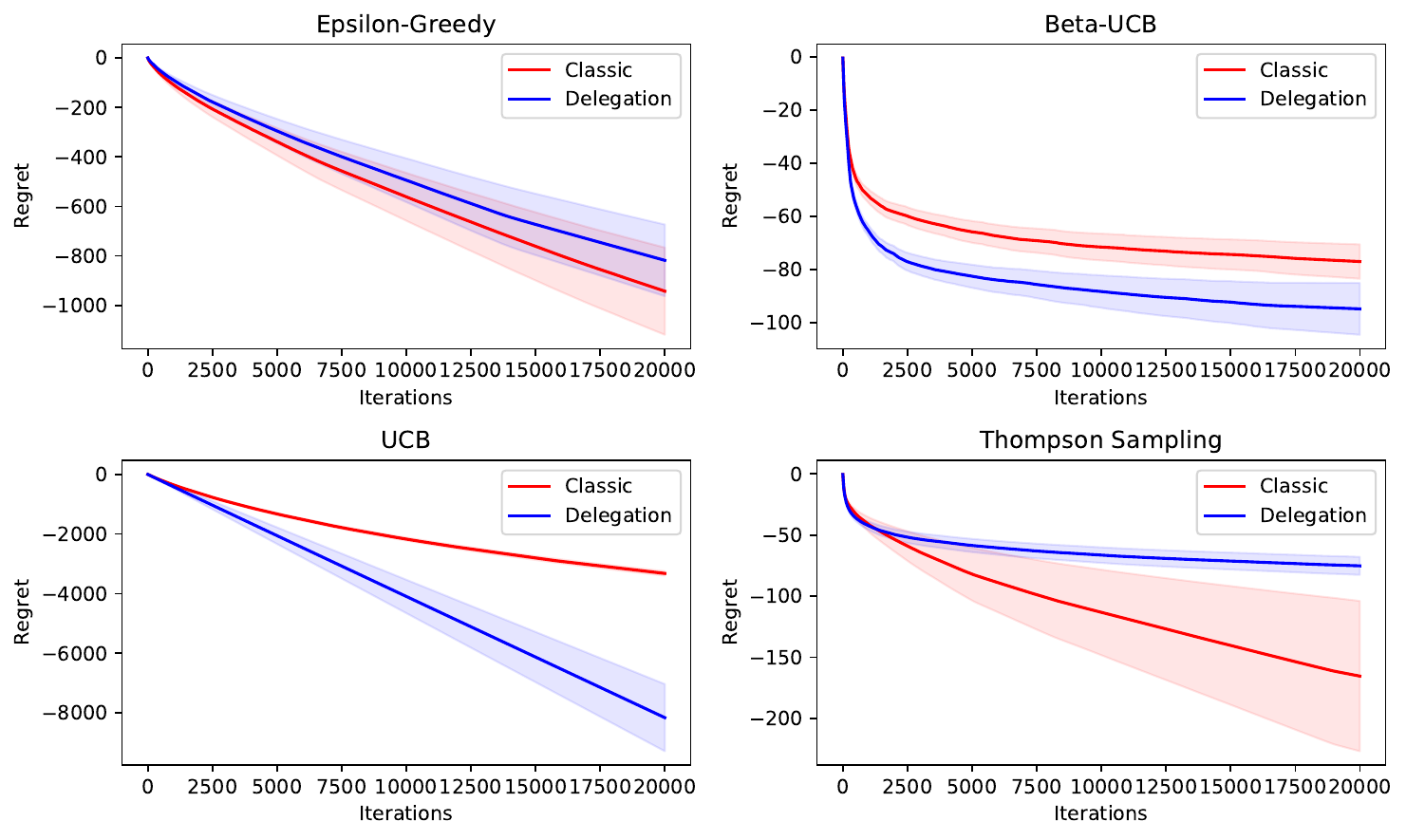}
\caption{\label{fig:eval5} Evaluation of the four algorithms with 20 agents using a binomial graph topology with connectedness set to 0.6.}
\end{figure}

Turning our attention to the level of connectedness of the graph, Figure \ref{fig:eval5} illustrates the case where the probability of agent connection increases from 0.3 to 0.6. These results again illustrate that as the number of possible paths increases, the UCB algorithms struggle to cope with the additional exploration, while Thompson sampling still continues to perform well. We believe that this is due to the manner in which we propagate the results of Thompson sampling up the delegation chain which reduces the amount of exploration needed by the system.

\begin{figure}
\includegraphics[width=\textwidth]{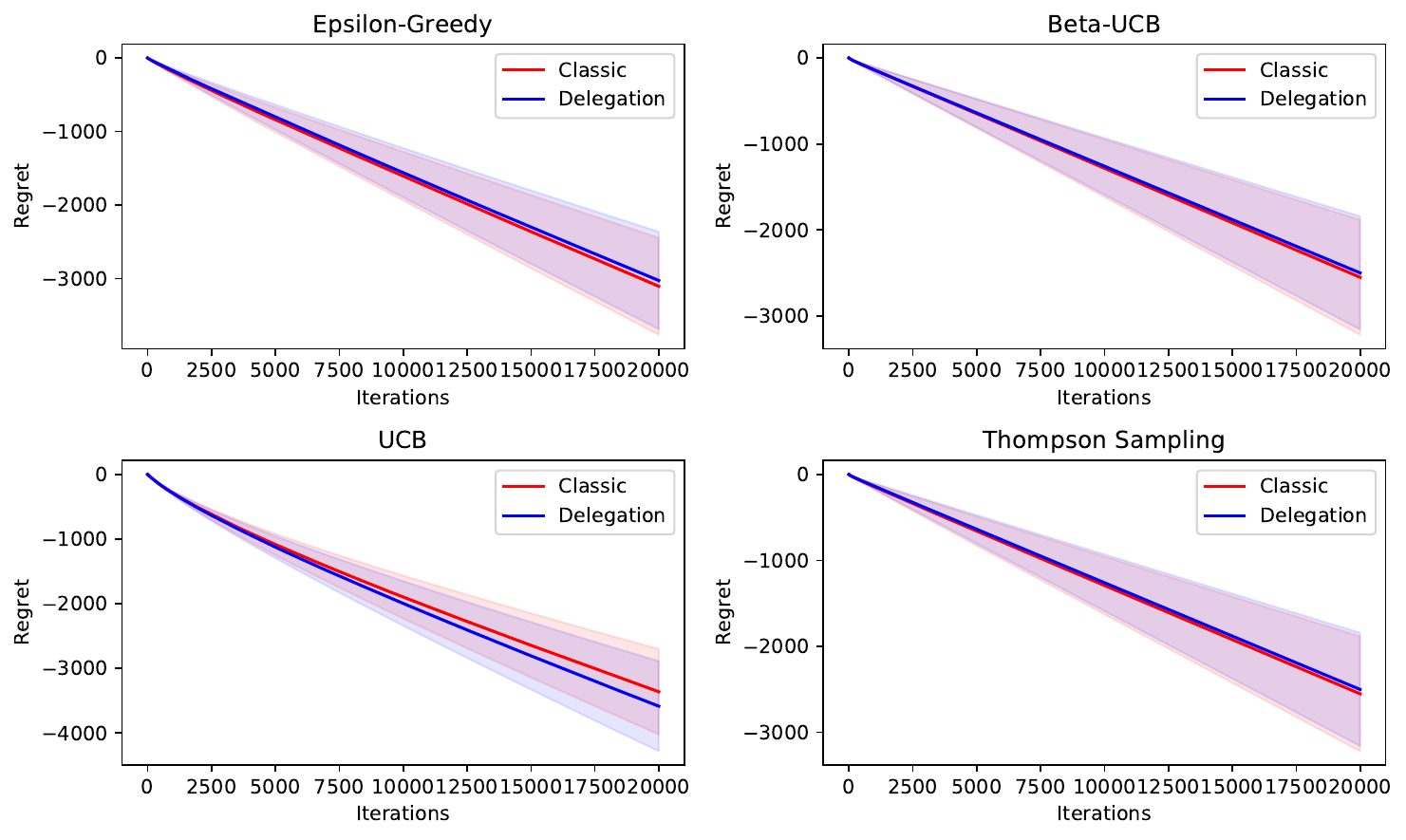}
\caption{\label{fig:eval6} Evaluation of the four algorithms over a small world topology.}
\end{figure}

Finally, we examined the impact of graph topology. Figure \ref{fig:eval6} shows the results when the algorithms are evaluated over a directed scale-free network. Here, the performance of the standard and variant algorithms is near-identical due to the low number of delegation choices available to the agents; this result shows how (unsurprisingly) our variant algorithms operate in a way similar to the classic algorithms in simple settings.

\section{Related and Future Work} \label{sec:discussion}

As mentioned above, while there has been a significant amount of work in the area of trust and reputation systems, relatively little research has considered the problem of recursive delegation. Afanador et al. \cite{afanador2019algorithms,afanador19phd} tackled the problem through the use of \emph{delegation games}. Afanador et al. also considered another multi-arm bandit based approach to recursive delegation, namely through an approximation of the Gittens index, to determine which agent to delegate to.

Another examination of recursive delegation was undertaken by Burnett and Oren \cite{burnett2012sub}. In this work, the authors examined how rewards and punishments (in the form of increasing/decreasing trust values) should be apportioned to agents depending on their position in a delegation chain.

The goal of this paper was to highlight the recursive delegation problem noting that standard algorithms often work poorly in this domain, and to encourage additional research in the area. While one important avenue of future work is to identify additional effective algorithms to cater for recursive delegation, multiple additional pathways for future research exist, which we intend to pursue. For example, we note that in the current formalisation, agents have full visibility of the system. In realistic scenarios, delegators and delegatees may have partial information about different parts of the domain. This could include knowledge of the topology of the delegation graph and partial information about who was delegated to. An example of the latter is that when a task is executed successfully, a delegatee may wish to claim credit without informing the delegator about who executed the task, while if failure happens, the delegatee may wish to blame who they delegated to for the failure, introducing mitigating circumstances \cite{DBLP:conf/atal/MilesG15a}. Other variants of the problem could involve new agents entering and exiting the system; agents being busy with other tasks; costs associated with delegations (which would potentially relate the problem to principal agent theory \cite{LaffMart01}); the incorporation of indirect trust information into the system; and so on. In addition, the large body of work covering trust and reputation systems has considered many different facets of the (single) delegation problem, and many of these could need reconsideration in the context of recursive delegation.

\section{Conclusions}
In this paper we considered the problem of recursive delegation, modelling it as a (recursive) multi-arm bandit problem. We modified the $\epsilon$-greedy; Thompson sampling and two versions of the UCB algorithms to cater for recursive delegation, and evaluated these modified version against their standard counterparts across different hyperparameters; number of agents and graph topologies. Our results indicate that in many situations the recursive delegation aware variants (particularly Thompson sampling) outperform the standard techniques, though the additional exploration inherent in the recursive delegation case can significantly affect algorithm performance.

Our work seeks to highlight the recursive delegation problem and its importance to real-world settings while aiming to spur additional research in the area. This paper serves as a first step to investigations in the domain, and we have highlighted several important avenues for future work.


\bibliographystyle{abbrv}
\bibliography{main}
\end{document}